\documentclass[12pt]{article}
\usepackage{amsmath,amsthm,amssymb,amscd,parskip}

\usepackage[dvipdfm]{graphicx}
\usepackage[dvips]{color}
\usepackage{calc}
\usepackage{caption}
\usepackage{subcaption}
\usepackage{graphicx}
\usepackage{lscape}
\usepackage[nolists,tablesfirst]{endfloat}

\def\bSig\mathbf{\Sigma}

\renewcommand{\baselinestretch}{1.5}
\usepackage[figuresright]{rotating}
\usepackage{rotating}

\begin{document}

\title{Evaluating hospital infection control measures for antimicrobial-resistant pathogens
using stochastic transmission models: application to Vancomycin-Resistant Enterococci
in intensive care units}


\author{\small Yinghui Wei$^{1}$\footnote{Corresponding author: Yinghui Wei, Centre for Mathematical Sciences, Plymouth University, PL4 8AA, UK. Email: yinghui.wei@plymouth.ac.uk.}, Theodore Kypraios$^{2}$, Philip D. O'Neill$^{2}$,\\
\small Susan S. Huang$^{3}$,  Sheryl L. Rifas-Shiman$^{4}$ and Ben S. Cooper$^{5,6}$\\
\small $^{1}$ Centre for Mathematical Sciences, Plymouth University, UK\\
\small $^{2}$School of Mathematical Sciences, University of Nottingham, UK\\
\small $^{3}$ Division of Infectious Disease and Health Policy Research Institute,\\
\small University of California Irvine School of Medicine, Irvine, California, USA\\
\small $^{4}$Department of Population Medicine, Harvard Medical School \\
\small and Harvard Pilgrim Health Care Institute, Boston, Massachusetts, USA\\
\small $^{5}$ Mahidol Oxford Tropical Medicine Research Unit (MORU), Bangkok, Thailand \\
\small $^{6}$  Centre for Tropical Medicine and Global Health, Nuffield Department of Clinical Medicine,\\
\small University of Oxford, Oxford, UK}

\date{}

\maketitle

\label{firstpage}

\begin{abstract}
Nosocomial pathogens such as Methicillin-Resistant {\em Staphylococcus aureus} (MRSA) and
Vancomycin-resistant {\em Enterococci} (VRE) are the cause of significant morbidity and mortality among
hospital patients. It is important to be able to assess the efficacy of control measures using
data on patient outcomes. In this paper we describe methods for analysing such data using
patient-level stochastic models which seek to describe the underlying unobserved process of
transmission. The methods are applied to detailed longitudinal patient-level
data on VRE from a study in a US hospital with eight intensive care units (ICUs). The data comprise admission
and discharge dates, dates and results of screening tests, and dates during which precautionary measures
were in place for each patient during the study period. Results include estimates of the efficacy
of the control measures, the proportion of unobserved patients colonized with VRE and the proportion
of patients colonized on admission.
\end{abstract}

keywords:
Antimicrobial resistance; Bayesian methods; Healthcare-associated infections; MRSA; VRE

\section{Introduction}
\label{s:intro}
The emergence of antimicrobial-resistant pathogens such as Methicillin-Resistant {\em Staphylococcus aureus} (MRSA) and
Vancomycin-resistant {\em Enterococci} (VRE) during the past two decades is a major clinical and
epidemiological problem throughout the world. Understanding these pathogens and assessing methods for
their control continues to be an area of considerable importance (see e.g. \cite{deangelis14, iosifidis13, lee13} and references therein). In this paper we describe methods for analysing longitudinal patient-level hospital
data on nosocomial pathogens. The methods involve defining stochastic models which
describe how the pathogen spreads between individual patients. Specifically, we build on
the methods of \cite{kypraios10} in which data on MRSA were analysed, and extend
this approach to data on VRE. The focus of the earlier paper was towards clinical results, and in particular there was no detailed
description of the statistical methods used. In contrast, here we fully describe these methods, and
also develop methods for model comparison and model assessment which were not in the earlier work.

The kind of data we consider typically consist of admission and discharge times for each patient on a hospital
ward, along with the times and results of screening tests for the pathogen in question. There may also be other
information, such as the times and types of infection control measures administered, or treatments.
Such data are characterized by two aspects, namely that they are highly
dependent (e.g. whether or not a patient is colonized with the pathogen in question depends on the status of
other patients) and that the process which generates the data, namely transmission, is unobserved. This in turn
implies that standard statistical methods for longitudinal data such as time-series methods or survival
analysis methods are not usually natural in this setting. Instead, modelling methods are more appropriate.

Various modelling approaches to analysing longitudinal epidemiological data on hospital pathogens have been developed,
typically applied to data on MRSA. These include fitting deterministic models (\cite{austin99}), maximum-likelihood methods applied to Markov chain models \cite{pelupessy02, bootsma07, drovandi08}, hidden Markov models \cite{mcbryde07}, and non-parametric approaches for Markov process models \cite{wolkewitz08}. All of these methods involve
simplifying assumptions of one kind or another which can be relaxed by using individual-level stochastic
models and a Bayesian framework. The latter approach, using Markov chain Monte Carlo (MCMC) methods, is taken in
\cite{forrester07, kypraios10} essentially by developing the ideas in \cite{gibson98, Oneill99}. This is the approach that we adopt in
this paper.
Since the underlying models essentially
describe the unobserved process of transmission between patients, the
MCMC methods involve augmenting the observed data with the transmission
events themselves. The appeal of this approach is that it is highly flexible,
it usually avoids the need for simplifying assumptions which are common in other
methods, and can provide much richer inferential information from the data
than maximum likelihood methods.

In this paper we analyse data on VRE, which are bacteria from the genus {\em Enterococcus} which have
acquired resistance to Vancomycin. {\em Enterococci} also have intrinsic resistance to several classes of antibiotics.
They usually live harmlessly within the gastrointestinal tract or on the skin,
but can cause infections, especially among those with weakened
immune systems. In consequence, they can produce disease in highly vulnerable hospital
patients, such as those in oncology wards or intensive care units  (ICUs).
Changes in resistance patterns over time frequently
mean that VRE strains are resistant to most antimicrobial agents.
VRE have been endemic in hospital settings in Europe and the USA since the start of
the 21st century, and are a major problem in healthcare settings throughout the world.
In such environments, transmission of VRE between patients is believed to occur indirectly, largely via health care workers
and in particular by contamination of hands or gloves \cite{boyce95, bonten96}. Environmental contamination
is also thought to be a possible transmission route. For this reason, control strategies typically focus on activities such as hand-washing, barrier
precautions (glove and gown use), and environmental cleaning.

The paper is structured as follows. The data structure and transmission models
are described in sections 2 and 3, respectively. The MCMC methods are introduced
in section 4 and applied to the VRE data in section 5. We finish with some
discussion in section 6.

\section[Data]{Data structure}\label{Data}

The individual patient data are taken from a 17-month longitudinal study in
a hospital in Boston, USA, involving over 8,000 patients who were admitted to
one of eight different ICUs during the study period.
The data consist of the dates on which each patient was admitted and discharged from
the ICU, the dates and results of any screening tests that took place,
and any dates during which the patient was placed under precautionary measures due to
their being supposed to be colonized with VRE. The precautionary measures consisted of
glove and gown use by healthcare workers, and the use of isolation rooms, and were the
same for all patients concerned. Such data contain many inherent
dependencies: for instance, whether or not a patient becomes colonized could reasonably
be supposed to depend on the number of other currently-colonized patients in the ICU.
For this reason it is natural to analyse such data using a model-based approach which
explicitly describes how the pathogen can spread between patients over time.

Each patient may undergo a number (possibly zero for some patients) of screening tests,
the results of which are either positive (denoting colonization with the pathogen) or negative. Note that we take {\em colonization} to mean either asymptomatic carriage (i.e. the pathogen is present in the body but has not caused clinical infection) or infection, and we do not distinguish between these two states. Furthermore, for VRE the amount of pathogen shed by carriers or those infected is usually similar, which
further motivates this modelling assumption. Each patient may also be placed in isolation during their stay in the ward, usually as a result of
a positive test, or for other clinical reasons. The exact meaning of isolation depends on the study in question,
examples including physical isolation in a single room, use of additional barrier precautions, etc.
The models we describe below assume that only one kind of isolation is adopted, although it would be
straightforward to relax this assumption.

\section{Models}\label{Model}

The data set of interest comprises eight different ICU wards. We analyse the data from
each ICU separately. This is motivated by the fact that the ICUs have different kinds of patients
(e.g. medical/surgical) and so there is no reason to suppose that parameters
governing VRE transmission are common across all wards. Furthermore, relatively few
patients were admitted to more than one of the ICUs.

In the following, we describe modelling approaches for data on patients in a single ICU.
The study period begins and ends at times $T_S$ and $T_E$, respectively.
During the study period, the admission and discharge times of
all patients on the ward are recorded. Patients already present in the ward at time $T_S$ are
deemed to have been admitted at $T_S$, and those in the ward at $T_E$ are deemed to have been
discharged at $T_E$. Patients might be re-admitted during the study period.

We consider three models for colonization, with the following common assumptions.
First, at each point in time every patient in the ward is either susceptible or colonized,
and can be either in or out of isolation.
Second, the test results are assumed to have perfect specificity (so that a positive
test result can only arise if the patient really is colonized), and sensitivity
$p \times 100 \%$ (so a colonized patient has probability $p$ of testing positive).
The assumption of perfect specificity is natural in the setting we consider, since screening
tests are usually culture-based, and false positives can only arise via the rare event of contamination
\cite{perry04}. However, this assumption can be easily relaxed within our framework.

Third, if a patient has a positive test, then they are assumed to remain colonized
for the following six months \cite{huang07, robicsek09, scanvic00}.
In particular, if a patient is re-admitted within
six months of a positive test, then they are said to be {\em colonized on
re-admission} at the time of re-admission, and assumed to remain colonized
until next discharged. Conversely, if a patient is admitted to the ward
with no positive test result during the previous six months, which in many
cases is simply because they have never been previously admitted at all, then
they are called a {\em new admission} at the time of admission. Each new admission
has a probability $\phi$ of being {\em colonized on admission}, independently
of all other patients and admissions. We refer to $\phi$ as the {\em importation
probability}. New admissions are also formally regarded in the model as being
new patients, even if they have been previously
admitted. In particular, the term {\em patient} should be interpreted in this
way in this section and the next. Although the assumption that the same individual
can be formally treated as two patients is made largely for simplicity, it is often
pragmatic in practice unless there are a large number of re-admissions in the data.

The final component of the model is the mechanism for potential colonization of patients
who are not already colonized. We adopt the usual
assumption from stochastic disease transmission modelling that susceptible individuals
are colonized according to a Poisson process whose rate might depend on the numbers of
existing colonized patients within the ward. In the context of nosocomial pathogens, the
transmission between patients is usually indirect, typically via healthcare workers or the
environment. We refer to patients who are colonized in this way as {\em colonized on the ward}.

Specifically, each susceptible patient is, independently of other susceptibles,
assumed to be colonized at a time corresponding to the first point of a non-homogeneous Poisson process
of rate $\lambda(t) \geq 0$ at time $t$.  We consider three possible models for $\lambda(t)$.
In order to investigate the effectiveness of control measures, in each model we
differentiate between isolated and non-isolated patients.

{\em Full model.} This model is defined in \cite{forrester07} and assumes that
\[
\lambda(t) = \beta_0 + \beta_1 C(t) + \beta_2 Q(t)
\]
where $Q(t)$ and $C(t)$ denote the number of colonized patients on the ward at time $t$ who
are in and out of isolation, respectively. Here $\beta_0$ represents a background
rate of colonization which is unaffected by the current prevalence on the ward, for instance
arising due to medical staff who service many wards and
may act as vectors for the pathogen. Conversely,
$\beta_1$ and $\beta_2$ represent rates attributed to colonized patients. This way of
modelling transmission is a natural generalization of the usual mass-action assumption in
standard epidemic models such as the SIR (susceptible-infective-removed) epidemic model
(see e.g. \cite{Oneill99}), which has itself been extensively used to successfully describe
pathogen transmission in many different settings. Specifically, our approach assumes that
each colonized individual with the same isolation status (in or out) makes an equal and
independent contribution to the overall colonization rate. This is a very natural assumption
with a clear epidemiological interpretation, and also avoids the need to specify more complex
dependencies between individuals.

{\em No-background transmission model.} This model is a special case of the full model in which $\beta_0$
is assumed to be close to zero with high probability. In practice this is achieved by
assigning a suitable prior distribution to $\beta_0$, as discussed below. This model represents
the hypothesis that transmission almost always occurs due to the presence of colonized patients
in the ward. As well as being of interest in its own right, considering this model also allows us to
assess the impact of assuming background transmission.

{\em Non-linear model.} This model assumes that it is only the presence of colonized patients,
not their number, which affects the rate of new colonizations. In particular it is not
assumed that $\lambda(t)$ is linear in $C(t)$ and $Q(t)$, as in the full model. Such a
model is suitable if the amount of pathogen in the ward from one patient is so great as to
have a saturation effect. Specifically,
\[
\lambda(t) = \beta_0 + \beta_1 \chi_{ \left\{ C(t) > 0 \right\}} + \beta_2 \chi_{ \left\{ Q(t) > 0 \right\}},
\]
where $\chi_A$ denotes the indicator function of the event $A$.  In some sense this model provides an
extreme alternative to the mass-action type assumptions in the previous models, since now there is no
increase in colonization rate as the numbers of colonized individuals increase beyond one.

\section{Inference methods}
\subsection{Notation and likelihoods}

Let $\theta = (p, \phi, \beta_0, \beta_1, \beta_2)$ denote the vector of model parameters.
We denote all data (admission and discharge times, times and outcomes of test results, times of patient isolation) by $y$,
although as explained above we only have a probability model for the colonization process and test sensitivity.
Adopting a Bayesian framework, our objective is to explore the posterior density $\pi(\theta | y)$.
Now the likelihood $\pi(y | \theta)$ is intractable in practice, since its evaluation involves
integrating over all possible unobserved colonization times (cf. \cite{Oneill99}, for
the analagous situation for standard epidemic models). Consequently, we augment the parameter
space to include all colonization times, $c$ say.

Under the assumption of perfect test specificity,
$c$ includes the unobserved colonization time for each patient who ever has a positive test
result. Additionally, individuals who do not have a positive test result may still be colonized,
their undetected status arising either because they never had a test at all, or because
any test was a false negative. The colonization
times of such individuals are also included in $c$. Note also that $c$ is also assumed to describe
which patients are colonized on admission and which are colonized on re-admission.

By Bayes' Theorem,
\[
\pi( \theta, c | y) \propto \pi(y, c | \theta) \pi(\theta),
\]
where $\pi(\theta)$ denotes the prior density of $\theta$. It remains to evaluate the augmented
likelihood $\pi(y,c | \theta)$.
Let $n_A$ denote the number of new admissions and $n_{CA}$ denote the number of these
who are colonized on admission. Note that $n_A$ is determined by the data $y$, but $n_{CA}$ is
determined by both $y$ and $c$.
Define $n_{TP}$ and $n_{FN}$ as the number of true positive and false negative test results.
Under the assumption of perfect specificity, $n_{TP}$ is determined by $y$. Conversely,
since sensitivity need not be perfect, $n_{FN}$ is determined by $y$ and $c$.

Define $\mathcal{K}$ as the set of patients who are colonized on the ward.
For a typical patient in $\mathcal{K}$, denoted $j$ say, define $n_C(j)$
and $n_Q(j)$ as the numbers of patients who are respectively colonized and not isolated, and colonized
and isolated, at the time of $j$'s own colonization. Note that $\mathcal{K}$, $n_C(j)$
and $n_Q(j)$ can all be determined from $y$ and $c$, but none are known explicitly from just $y$.

The augmented likelihood for the full model takes the form
\begin{eqnarray}\label{full model}
\pi(y,c|\theta) &\propto& \phi^{n_{CA}} (1-\phi)^{n_A-n_{CA}} \times p^{n_{TP}}(1-p)^{n_{FN}}\nonumber\\
&\times& \prod_{j\in \mathcal{K}} \left(\beta_0 + n_C(j) \beta_1 + n_Q(j) \beta_2\right)\nonumber\\
&\times& \exp\left\{-\int_{T_S}^{T_E}{\Large(\beta_0S(t)+\beta_1C(t)S(t)+
\beta_2Q(t)S(t) \Large) \; dt}\right\}, \label{like}
\end{eqnarray}
where $S(t)$ denotes the number of susceptible patients on the ward at time $t$. The likelihood can be derived as follows.
First, the terms involving $\phi$ account for the probability of $n_{CA}$ patients being colonized on admission and the remaining $n_A - n_{CA}$
being not colonized on admission. Second, the terms involving $p$ give the probability of observing the true positive and false negative
test results. Note that since we keep track of each individual patient, for both the $\phi$ and $p$ expressions we do not require any
combinatorial terms; for instance, we know exactly which patients are colonized on admission, not just the number.

The remaining part of the likelihood arises from the colonization process. Recall that if a non-homogeneous Poisson process with
rate $\xi(t)$ at time $t$ is observed during a time interval $[0,T]$, and $m$ points occur at times $0 \leq t_1 < t_2 < \ldots < t_m \leq T$, the likelihood is given by
\begin{equation}
\label{poissproclike}
\left( \prod_{i=1}^m \xi( t_i - ) \right) \exp \left\{ - \int_0^T \xi(t) \; dt \right\} ,
\end{equation}
where $\xi( t_i - )$ denotes the rate just before time $t_i$, see for example \cite{daley03}, Chapter 7.
The product term accounts for the points occurring and the exponential term accounts for the absence of points at other times.
In our setting, each susceptible patient is colonized according to a Poisson process of rate
$\lambda(t) = \beta_0 + \beta_1 C(t) + \beta_2 Q(t)$ at time $t \geq 0$, and so the overall colonization process (i.e. the process
which counts colonization events but does not specify which individuals are actually colonized) is a Poisson process of rate
$S(t) \lambda(t)$. Setting $\xi (t) = S(t) \lambda(t)$ in (\ref{poissproclike}) gives the likelihood of this overall colonization process.
However, we also know the identity of each colonized patient, and since each susceptible patient is equally likely to be colonized at any
given time, the probability that a particular patient is colonized is $1/S(t)$. Incorporating this probability into the product term in
(\ref{poissproclike}) yields the product term in (\ref{like}). Finally, the likelihood in (\ref{like})
can also be derived and expressed by purely thinking in terms of individual patients, which results in the integral being expressed as
a sum over patients, see \cite{forrester07}.

It is straightforward to modify (\ref{like}) to obtain the
corresponding likelihoods for the no-background and non-linear models.
Finally, we assign independent Beta prior distributions to $p$ and $\phi$, and
independent exponential prior distributions
to the colonization rate parameters $\beta_0$, $\beta_1$ and $\beta_2$.

\subsection{Markov Chain Monte Carlo algorithm}\label{MCMC}
In order to explore $\pi(\theta, c | y)$ we use a Metropolis-Hastings MCMC algorithm which
updates the elements of $\theta$ and the unknown colonization times $c$, as follows.
First, it follows from (\ref{like}) and our choice of prior distributions
that both $p$ and $\phi$ have Beta-distributed full conditional distributions, and
hence can be updated according to Gibbs steps. Second, the colonization rate parameters
$\beta_0$, $\beta_1$ and $\beta_2$ can all be updated using a Gaussian random walk,
where proposed negative values are rejected immediately since the proposed
likelihood is zero.

Third, updates for the colonization times $c$ are achieved via three possible steps,
one of which is chosen uniformly at random during each iteration of the MCMC algorithm.
Recall that patients who are colonized on re-admission are assumed to have a colonization
time equal to their time of admission. Since such re-admissions are determined directly
by the data, the colonization times in question are never updated in the MCMC algorithm.

Of the remaining patients, those with a positive test result are assumed to be definitely
colonized, and such patients therefore always have a colonization time. Denote the set
of such patients $\mathcal{P}$ and note that $\mathcal{P}$ is determined by the data $y$.
All other remaining patients may or may not be colonized; denote the set of those
currently colonized by $\mathcal{N}_1$ and those not by $\mathcal{N}_0$. Thus $\mathcal{P}$,
$\mathcal{N}_0$ and $\mathcal{N}_1$ have no intersection and their union is the set
of all patients other than those colonized on re-admission. Let $n_0$, $n_1$ and $n_p$ denote
the number of patients in $\mathcal{N}_0$, $\mathcal{N}_1$ and $\mathcal{P}$, respectively.

The three possible steps involved in updating the colonization times are as follows.
We define $a_j$, $d_j$ and $c_j$ as the admission, discharge and colonization times,
respectively, of individual $j$.
\begin{enumerate}
\item Adding a colonization time: an individual $j$ is selected randomly from
$\mathcal{N}_0$. With probability $\phi_0$, $j$ is proposed to be colonized on admission, so
$c_j = a_j$; otherwise $c_j$ is sampled
uniformly from $(a_j,d_j)$. Denoting $\tilde{c}$ as the proposed new $c$ with $c_j$
added, the
move is accepted with probability
\[
\frac{n_0(d_j-a_j) \pi( \theta, \tilde{c} | y )}{(1-\phi_0)(n_1+1)
\pi( \theta, c | y)} \wedge 1
\]
if $c_j\in(a_j,d_j)$, and
\[
\frac{n_0 \pi( \theta, \tilde{c} | y )}{\phi_0(n_1+1) \pi(\theta, c | y)} \wedge 1
\]
if $c_j=a_j$, where $x \wedge y$ denotes $\min(x,y)$.

\item Deleting a colonization time: an individual $j$ is randomly selected from
$\mathcal{N}_1$, and its colonization time $c_j$ is removed from
$c$ with probability
\[
\frac{(1-\phi_0)n_1 \pi(\theta, \tilde{c} | y)}
{(n_0+1)(d_j-a_j) \pi(\theta, c | y )} \wedge 1
\]
if $c_j\in(a_j,d_j)$, and
\[
\frac{\phi_0 n_1 \pi(\theta, \tilde{c} | y)}{(n_0+1) \pi(\theta, c | y ) } \wedge 1
\]
if $c_j=a_j$, where $\tilde{c}$ denotes $c$ with $c_j$ removed.

\item Moving a colonization time: randomly select an individual $j$
from $\mathcal{N}_1 \cup \mathcal{P}$.
With probability $\phi_0$, set the proposed new colonization time as $\tilde{c}_j = a_j$;
otherwise, $\tilde{c}_j$ is sampled uniformly from $(a_j,t_j)$,
where $t_j = d_j$ if $j \in \mathcal{N}_1$
and $t_j$ is the time of $j$'s first positive test if $j \in \mathcal{P}$.
Thus the proposal density of $\tilde{c}_j$ is
\begin{eqnarray} \label{p1}
q( \tilde{c}_j | c_j)  = q( \tilde{c}_j) = \left\{
\begin{array}{ll} \frac{\phi_0}{n_1+n_p} & \mbox{if $\tilde{c}_j = a_j$},\\
\frac{1-\phi_0}{(n_1+n_p)(t_j-a_j)} & \mbox{if $\tilde{c}_j \in(a_j,t_j)$}.
\end{array} \right. \nonumber
\end{eqnarray}
The new colonization time $\tilde{c}_j$ is accepted with probability
\[
\frac{\pi(\theta, \tilde{c} | y) q( c_j | \tilde{c}_j)}
{\pi( \theta, c | y)q( \tilde{c}_j | c_j) } \wedge 1,
\]
where $\tilde{c}$ denotes $c$ with $c_j$ replaced by $\tilde{c}_j$.

\end{enumerate}

\subsection{Model assessment}
Although our main focus is not to formally distinguish between competing models,
it is nevertheless of interest to perform a comparison. There is no canonical
method to do this in our setting, and different approaches have their own pros
and cons. For simplicity, we used a form of
deviance information criterion (DIC, \cite{spiegelhalter02}).
Since our setting involves missing data (namely, the unobserved colonization times)
we require a form of DIC which takes this into account, and moreover which can
be readily calculated. We therefore used DIC$_6$ from \cite{celeus06} defined by
\begin{equation}
\mbox{DIC}_6 = -4E_{\theta,c}[\log  \pi (y,c|\theta)]
+ 2E_{c}[\log \pi (y,c | \hat{\theta}(y))| y, \hat{\theta}(y)],
\end{equation}
where $\pi(y,c | \theta)$ is given by (\ref{like}) and we set $\hat{\theta}(y)$ to be
the posterior mean of $\theta$, as estimated from our MCMC algorithm. In practice, DIC$_6$ can be easily
evaluated using MCMC output as described in \cite{celeus06}. Although we used
DIC$_6$ for convenience, it has been explored in a similar context to ours by \cite{worby13} who
found that it can be effective provided there are sufficient data available.

We also consider two forms of model assessment, i.e. methods for exploring how well each model
fitted the observed data. The first of these is a posterior predictive $p$-value \cite{gelman96}
to assess the model fit to data, specifically with the total number of detected colonizations as the
discrepancy statistic to measure the difference between the observed and simulated data. To implement
this method, the model parameters in question are first repeatedly sampled from their posterior distribution, i.e.
from the MCMC output. In practice we take every 100th sample from the chain. For each sampled set of parameter
values, the model is then simulated forward in time to produce a possible realisation. We then calculate a
$p$-value as the proportion of simulations in which the total number of detected colonizations is greater or equal to
that in the observed data. A $p$-value close to 0.5 indicates a good fit to the data. In contrast, an
extreme $p$-value ($p>0.95$ or $p<0.05$) indicates that the fitted model is not appropriate for the data.

The second method of model assessment is to consider the predictive distribution of the number of detected
colonizations through time. This is implemented by again sampling parameters from the posterior distribution
and simulating the model forward in time for each set of parameter values to produce a large number of realisations
of the whole process. We then graphically compare the observed data with the mean and quantiles of the set of simulated
realisations.

\section{Application to data}\label{Analysis}
\subsection{Data}
The data we consider were collected from eight 10-bed ICUs in a tertiary
academic medical center in Boston, Massachusetts over a 17-month period.
The ICUs comprised two medical wards (which we denote M1, M2), two general surgical wards (GS1, GS2), and four specialty surgery wards (SS1-SS4).
Routine rectal admission and weekly screening was carried out using swab tests, with compliance around 90\%.
In addition there was a protocol in which patients who were identified as being VRE-positive were placed under contact precautions which consisted of gown and glove use by healthcare workers, and use of single rooms.

\begin{figure}[t]
\begin{center}
\includegraphics[width=1cm\textwidth,height=16cm,width=8cm, angle=90]{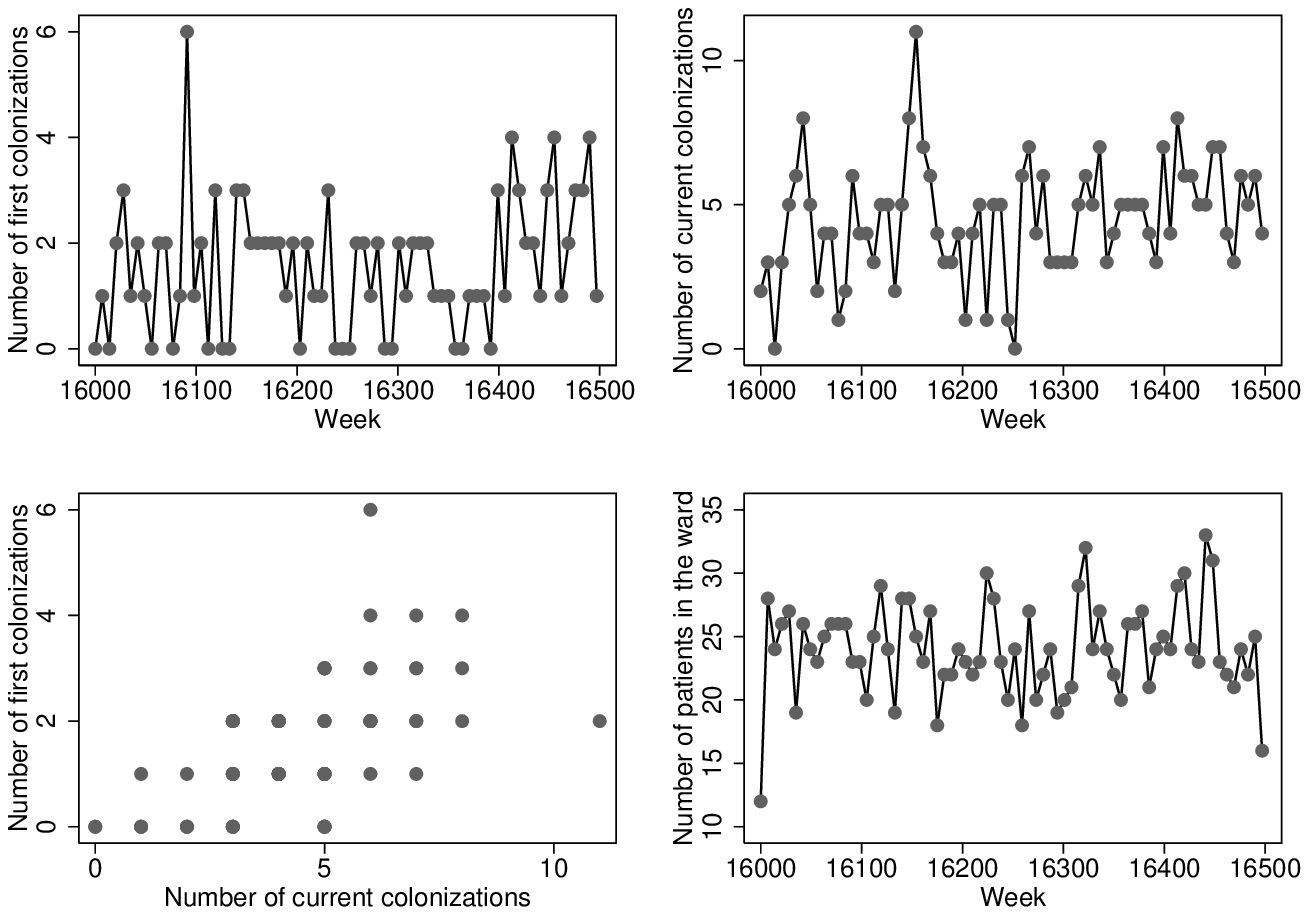}
\caption{Ward M1. Number of first detected colonizations, current total detected colonizations and number of patients on a weekly basis during the observation period. We also show the relationship between the number of first colonizations and number of current colonizations.} \label{Fig:explore_ward12D}
\end{center}
\end{figure}

Figure \ref{Fig:explore_ward12D} shows the number of first detected colonizations and total current detected colonizations for one of the wards (medical ward M1) on a weekly basis during the observation interval. Descriptive statistics of the individuals in ICUs are shown in Table \ref{t:summary statistics}. The relatively high values for the standard deviation of length of stay reflect the fact that
ICU patients typically have either quite short or quite long stays. The median length of stay
was around 3-6 days. On most wards, patients have just over one swab test on average, and thus
there are many patients in the data set with only one (usually admission) swab test. Finally, around 5-10\% of all
patients in the study were re-admitted during the study period.

\begin{table}[b]
 \vspace*{-6pt}
 \centering
 \def\~{\hphantom{0}}
  \caption{Summary statistics for VRE data.}
\label{t:summary statistics}
  \begin{tabular*}{\columnwidth}{@{}l@{\extracolsep{\fill}}c@{\extracolsep{\fill}}r@{\extracolsep{\fill}}r@{\extracolsep{\fill}}r@{\extracolsep{\fill}}r@{\extracolsep{\fill}}l@{\extracolsep{\fill}}c@{\extracolsep{\fill}}c@{\extracolsep{\fill}}c@{}}
  \hline
        & & & & Number & Number of \\ [1pt]
 & & &Percent& of swab& VRE+ \\
       &Number  & Length &in&tests per&  swab tests \\
    &of & of stay & contact&person&  per person \\
 Ward   &patients  & Mean(SD)& precautions& Mean(SD)& Mean(SD) \\

\hline
M1& $1293$ & $3.4(4.7)$&$11.4$&$1.2(0.8)$ & $0.2(0.8)$\\
M2& $1018$ & $4.4(6.4)$&$19.1$&$1.4(1.1)$ & $0.7(1.6)$\\
GS1& $1227$ & $3.4(5.2)$&$12.4$&$1.1(0.9)$ & $0.3(0.9)$\\
GS2& $1030$ & $4.0(8.3)$&$10.7$&$1.0(1.3)$ & $0.1(0.7)$\\
SS1& $706$ & $5.8(11.4)$&$12.5$&$1.4(1.6)$ & $0.7(2.2)$\\
SS2& $888$ & $4.9(9.7)$&$7.5$&$1.3(1.4)$ & $0.3(1.8)$\\
SS3& $1097$ & $3.8(6.4)$&$6.0$&$1.3(0.9)$ & $0.1(0.6)$\\
SS4& $1263$ & $3.6(5.2)$&$5.1$&$1.1(1.0)$ & $0.1(0.5)$\\
\hline
\end{tabular*}\vskip18pt
\end{table}


\subsection{Algorithm implementation}
Before considering the data at hand, we first simulated data
sets from each of the true models and then used our MCMC algorithm to estimate
the model parameters. The results indicated that our methods could recover the true
parameter values reasonably well.
The MCMC algorithm was then implemented separately
for the eight wards for each of the three models under consideration,
and the DIC values computed. The model parameters were assigned uninformative
independent prior distributions. Specifically, $\phi, p \sim U(0,1)$ and
$\beta_0, \beta_1, \beta_2 \sim Exp(10^{-6})$ where $Exp(\lambda)$ denotes
an exponential distribution with mean $\lambda^{-1}$, with the exception
that $\beta_0 \sim Exp(10^6)$ in the no-background model. The algorithm
was implemented by using Microsoft Visual C++ 2010 Express software with
double precision and the GSL Scientific Library. The C++ code is
available upon request from the first author.

\subsection{Results}
{\em Parameter estimates.} Posterior medians of the model parameters
are shown in Table \ref{t:posterior}. Estimates of both the importation probability
$\phi$ and the test sensitivity $p$ appear robust to the choice
of model, and both parameters are estimated with reasonable precision.
The importation probability is generally in the range 5-15\% other than
medical ward M2 where it exceeds 20\%. The test sensitivity is
around 70\% or higher, with the striking exception of general surgery ward
GS2, where it is around 50\%. These values for $\phi$ and $p$ are both
reasonable in clinical terms.

The transmission rate parameters estimates exhibit clear variation between
the models, which is to be expected due to the differences of the model
assumptions themselves. The posterior standard deviations are typically
of the same order as the posterior median values, illustrating that
there is reasonable uncertainty in the posterior estimation of $\beta_0$,
$\beta_1$ and $\beta_2$. This uncertainty is essentially a consequence of
the prevalence of VRE (discussed below): the majority of patients never
have a positive test which makes it harder to estimate transmission rate
parameters accurately.

The background rate $\beta_0$ is (in the full and non-linear models) invariably
estimated to be higher than $\beta_1$ and $\beta_2$, often around twice as much or more.
This can be interpeted as saying that the background colonization rate is roughly
equivalent to that due to two or three colonized patients on the ward. Regarding estimation
of such rates, note that any colonization event which occurs whilst the ward
contains no colonized patients must be attributable to the background rate, whereas
colonization events that occur with (as is typical) one or two colonized patients
in the ward could be due to background or transmission events. Thus it is possible
that, if we explicitly modelled the `source' of each colonization event, more
colonizations than not would be background colonization events.

\renewcommand{\baselinestretch}{1.0}

\begin{table*}
 \centering
 \def\~{\hphantom{0}}
\caption{Posterior median and standard deviation
 of all model parameters.}
 \label{t:posterior}
  \centering
\begin{tabular*}{\columnwidth}{@{}l
@{\extracolsep{\fill}}c@{\extracolsep{\fill}}c@{\extracolsep{\fill}}c@{\extracolsep{\fill}}c@{\extracolsep{\fill}}c
}
 \hline\\[-6pt]
     Ward   & $\phi$&$p$ &$\beta_0$& $\beta_1$& $\beta_2$
            \\[1pt]
\cline{2-6} \\[-6pt]
 &   \multicolumn{5}{c}{Full model}\\[1pt]
        \cline{2-6} \\[-6pt]
M1	&0.12	(0.02) &0.78 (0.03) &0.0084 (0.004) &0.0023 (0.002) &0.0025 (0.003)\\
M2	&0.23	(0.02) &0.81 (0.02) &0.0093 (0.006) &0.0028 (0.002) &0.0029 (0.002)\\
GS1	&0.12	(0.01) &0.78 (0.03) &0.0075	(0.005) &0.0057 (0.004) &0.0034 (0.003)\\
GS2	&0.05	(0.01) &0.49 (0.05) &0.0082	(0.005) &0.0034 (0.003) &0.0037 (0.003)\\
SS1	&0.13	(0.02) &0.84 (0.02) &0.0038	(0.004) &0.0014 (0.002) &0.0050 (0.003)\\
SS2	&0.05	(0.02) &0.71 (0.05) &0.0088	(0.006) &0.0065 (0.004) &0.0059 (0.004)\\
SS3	&0.10	(0.02) &0.66 (0.05) &0.0067	(0.004) &0.0029 (0.003) &0.0048 (0.004)\\
SS4	&0.04	(0.01) &0.68 (0.06) &0.0042	(0.002) &0.0023 (0.003) &0.0069 (0.006)\\
\hline\\[-6pt]
& \multicolumn{5}{c}{No-background model}\\[1pt]
 \cline{2-6} \\[-6pt]
M1	&0.13   (0.01)&0.77 (0.03) &$10^{-4}$  	$(2 \times 10^{-4})$ &0.0066 (0.003) &0.0054   (0.003)\\
M2	&0.23   (0.02)&0.81 (0.02) &$10^{-4}$	$(1 \times 10^{-4})$  &0.0050 (0.003) &0.0056   (0.003)\\
GS1	&0.13   (0.02)&0.77	(0.03) &$10^{-4}$	$(1 \times 10^{-4})$  &0.0099 (0.003) &0.0052   (0.003)\\
GS2	&0.05	(0.01)&0.49	(0.05) &$10^{-4}$	$(1 \times 10^{-4})$  &0.0071 (0.004) &0.0079	 (0.003)\\
SS1	&0.13	(0.02)&0.84	(0.02) &$10^{-4}$	$(1 \times 10^{-4})$  &0.0015 (0.002) &0.0069   (0.002)\\
SS2	&0.05	(0.02)&0.71	(0.04) &$10^{-4}$	$(1 \times 10^{-4})$ &0.0087 (0.004) &0.0104	 (0.005)\\
SS3	&0.11	(0.02)&0.65	(0.05) &$10^{-4}$	$(1 \times 10^{-4})$ &0.0072 (0.003) &0.0060   (0.004)\\
SS4	&0.05	(0.01)&0.65	(0.06) &$10^{-4}$	$(2 \times 10^{-4})$ &0.0073 (0.004) &0.0074   (0.006)\\
\hline\\[-6pt]
& \multicolumn{5}{c}{Non-linear model}\\[1pt]
\cline{2-6} \\[-6pt]
M1	&0.12   (0.02) &0.77   (0.03) &0.0081 (0.004)   & 0.0040  (0.004)  &0.0040 (0.004)\\
M2	&0.23   (0.02) &0.81   (0.02) &0.0092 (0.006)   & 0.0050 (0.005)   &0.0071 (0.006)\\
GS1	&0.12	(0.01) &0.78   (0.03) &0.0113 (0.005)   & 0.0047 (0.005)   &0.0048  (0.005)\\
GS2	&0.05	(0.01) &0.49   (0.05) &0.0085 (0.005)   & 0.0061 (0.006)   &0.0039  (0.004)\\
SS1	&0.14	(0.02) &0.84   (0.02) &0.0048 (0.004)   & 0.0032 (0.004)   &0.0069  (0.005)\\
SS2 &0.05   (0.01) &0.71   (0.04) &0.0117 (0.007)   & 0.0081 (0.007)   &0.0052 (0.006)\\
SS3	&0.11	(0.02) &0.66   (0.05) &0.0071 (0.004)   & 0.0039 (0.004)   &0.0057 (0.005)\\
SS4	&0.04	(0.01) &0.68   (0.06) &0.0041 (0.002)   & 0.0028 (0.004)   &0.0035  (0.004)\\
\hline
\end{tabular*}
\end{table*}

\renewcommand{\baselinestretch}{1.5}

{\em Unobserved carriage.} Unobserved carriage arises in two ways, namely false negative test
results, or colonization which occurs before any test is carried out. Thus unobserved carriage
is a function of both the test sensitivity and the frequency of testing.
Our transmission models explicitly include all unobserved
colonization events, and so in particular it is possible for individuals to
be colonized but undetected. The inference methods enable us to estimate the
extent of this unobserved colonization directly. Table \ref{t:monthly prevalence}
shows the observed mean monthly prevalence of VRE, calculated directly from the positive
test results in the data, and the median posterior prevalence as predicted by the full model (other
models gave similar results). It appears that the observed prevalence accounts for
just over half of the predicted prevalence.

\begin{table}
\vspace*{-6pt}
\centering
\def\~{\hphantom{0}}
 \caption{Observed and predicted monthly VRE prevalence.}
\label{t:monthly prevalence}
\begin{tabular}{lll}
Ward & Observed & Predicted\\
& Mean (SD) & Median (SD)\\


\hline
M1  &  14.8  \% (2.9\%) & 23.3\% (3.7 \%)\\
M2  &  28.6  \% (5.2\%) & 41.3\% (5.0 \%)\\
GS1 &  18.2  \% (2.8\%) & 22.8\% (4.1 \%) \\
GS2 &  9.9   \% (2.9\%) & 15.6\% (4.0 \%)\\
SS1 &  18.9  \% (7.6\%) & 26.4\% (8.6 \%)\\
SS2 &  11.4  \% (6.5\%) & 16.9\% (5.5 \%)\\
SS3 &  9.0   \% (3.1\%) & 16.7\% (2.8 \%)\\
SS4 &  4.5   \% (3.0\%) & 6.9\%  (4.7 \%)\\
\hline
\end{tabular}\vskip18pt
\end{table}

It is possible to derive more complex quantities which describe the extent of
unobserved carriage. In particular, we define $P_{hidden}$ as the proportion of
all colonized-patient-days which are unobserved in the sense that a patient is
colonized but not under precautions, and $P_{wait}$ as the proportion of all
colonized-patient-days during which a patient is pending knowledge of a positive test result and has yet to be placed under precautions. Denote $p_j$ as the precaution time of individual $j$, with $p_j = \infty$ for an individual never placed under precautions, recall that $t_j$ is the time of $j$'s first positive test and set $t_j = \infty$ if $j$ never has a positive test. It follows that
\[
P_{hidden} = \frac{\sum_{j: c_j \in [a_j, d_j]} \left( \left[ (p_j \wedge d_j) - c_j \right] \vee 0 \right)}
{\sum_{j: c_j \in [a_j, d_j]} (d_j-c_j)}, \; \; P_{wait} = \frac{ \sum_{j: p_j \in [t_j, d_j]} (p_j-t_j)}{\sum_{j: c_j \in [a_j, d_j]} (d_j-c_j)},
\]
where $\wedge$ and $\vee$ denote $\min$ and $\max$, respectively. To see that $P_{wait}$ cannot exceed 1,
note that any $j$ in the numerator sum has $t_j < \infty$, and under the assumption of perfect specificity $a_j \leq c_j \leq t_j$, and thus $j$ also appears in the denominator sum.

\begin{figure}[t]
\begin{center}
\includegraphics[width=1cm\textwidth,height=8cm,width=16cm]{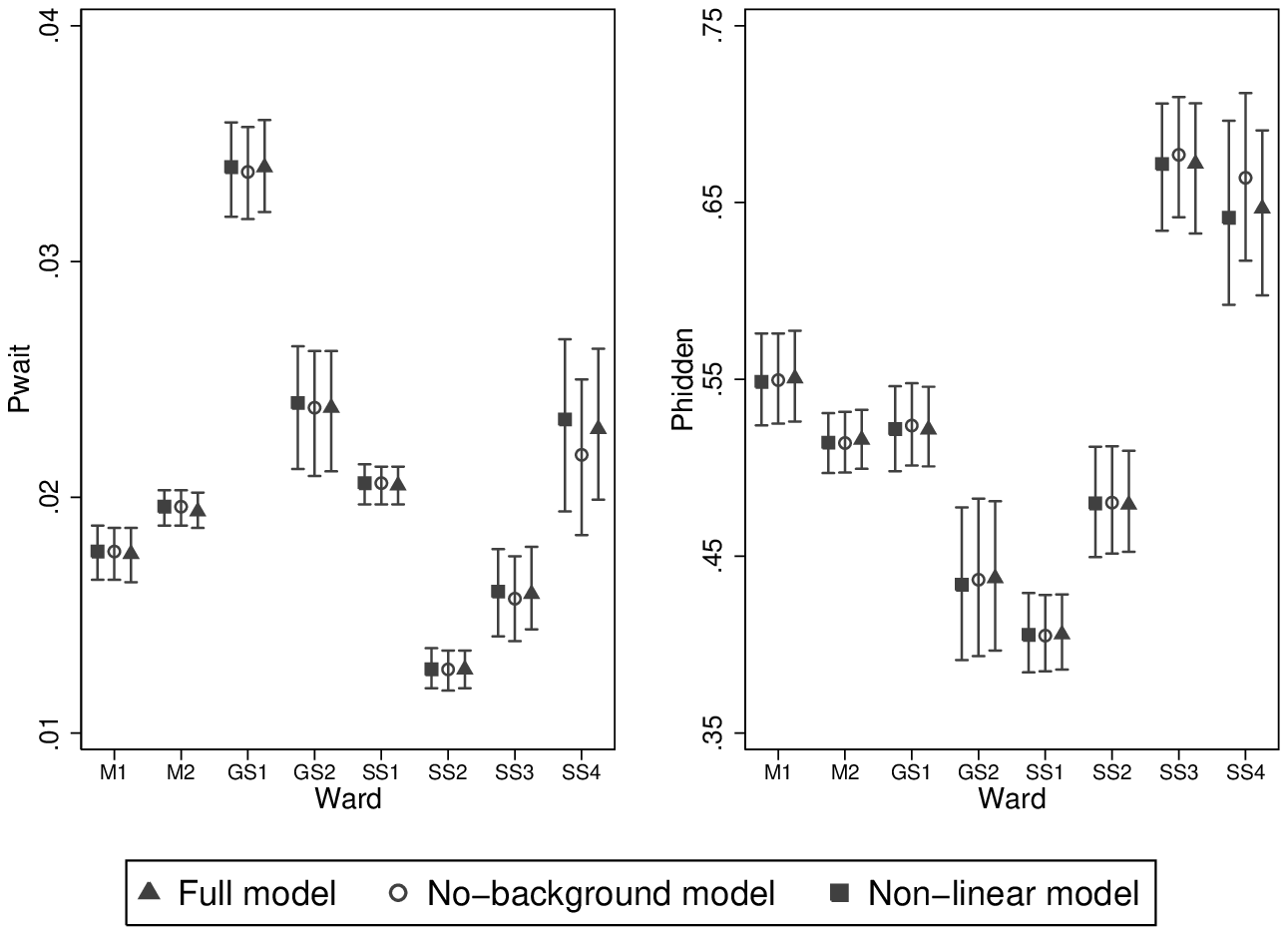}
\caption{Posterior median and $95\%$ credible intervals of the percentage of colonized-patient-days
 attributed to undetected colonized patients($P_{hidden}$), and the percentage of colonized-patient-days attributed to colonized patients who
 had been swabbed but were waiting for results ($P_{wait}$).} \label{Fig:PwaitandPhidden}
\end{center}
\end{figure}

Figure \ref{Fig:PwaitandPhidden} gives posterior median values of $P_{hidden}$ and $P_{wait}$ for all three models. The values appear to be robust across the models themselves. The apparently high values of $P_{hidden}$ are not surprising in view of the predicted prevalence values; in particular, each individual who is colonized but never placed under precautions contributes the entire duration of their colonization to the numerator of $P_{hidden}$. Conversely, the low $P_{wait}$ values indicate that patients who were detected were placed under precautions swiftly.

\begin{figure}[t]
\begin{center}
\includegraphics[width=1cm\textwidth,height=17cm,width=13cm]{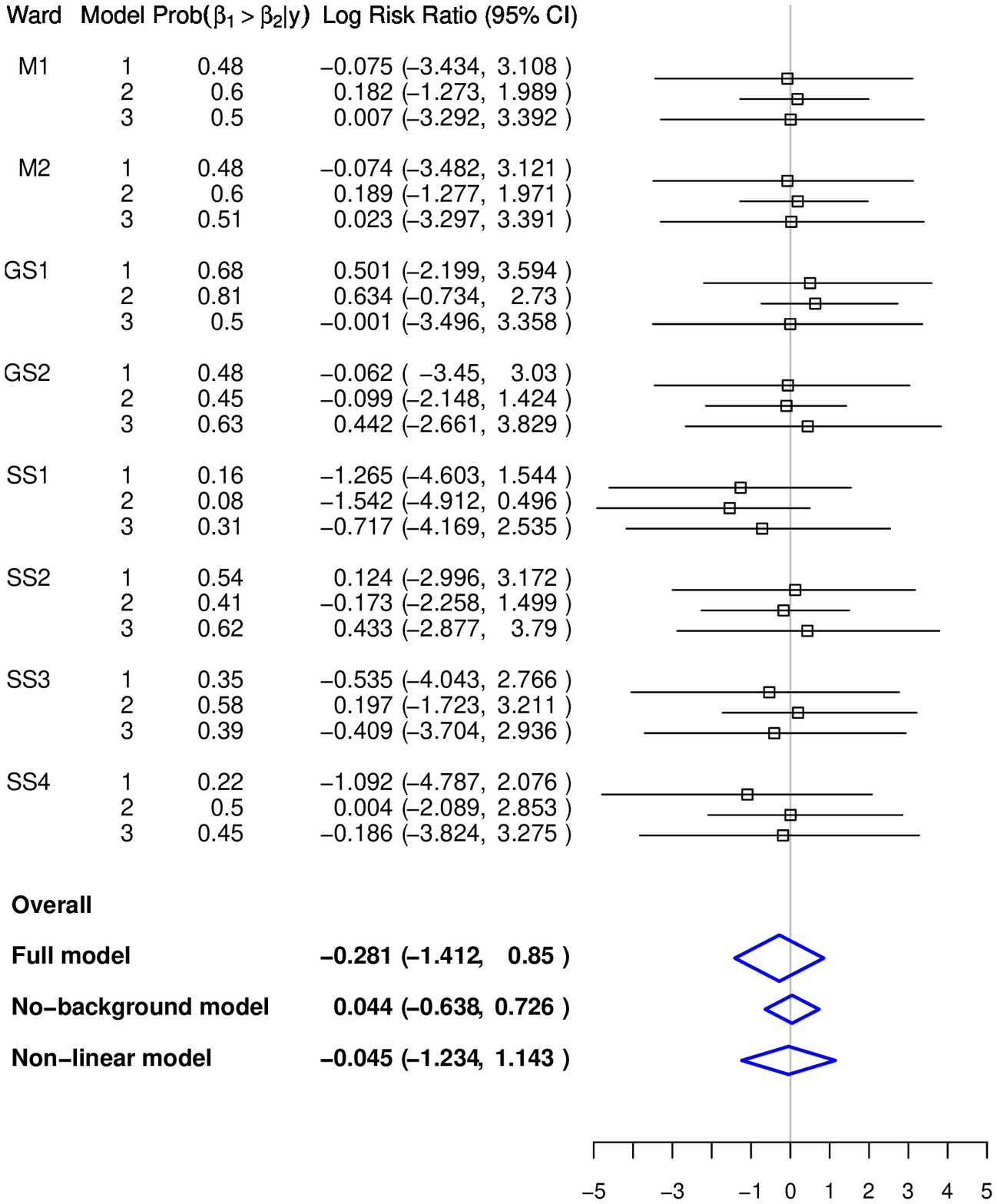}
\caption{Assessment of precaution measures efficacy via $P(\beta_1 > \beta_2|y)$
 and posterior median of $\log(\beta_1 / \beta_2$) with the $95\%$ credible intervals. The parameters $\beta_0, \beta_1$ and $\beta_2$ have independent exponential prior distributions with rate $10^{-6}$ . Each diamond box represents the overall precaution measures efficacy estimate from the meta-analysis and its $95\%$ confidence interval. Model indicators: 1 - Full model; 2 - No-background model; 3 - Non-linear model.} \label{Fig:efficacy}
\end{center}
\end{figure}

{\em Efficacy of precaution measures.}
The efficacy of the precaution measures in reducing transmission
may be evaluated by comparing the rate of $\beta_1$ and $\beta_2$,
with $\beta_1$ greater than $\beta_2$ indicating a positive benefit.
We specifically consider the posterior probability $P(\beta_1 > \beta_2 | y)$
and the posterior median value of $\beta_1 / \beta_2$, as given in
Figure \ref{Fig:efficacy}. From these values, there is no compelling
evidence to support the effectiveness of the precaution measures. The probability
$P(\beta_1 > \beta_2 | y)$ varies widely between wards, only once exceeding 0.8
(ward GS1) and with most values in the range 0.45-0.65. One exception is the
specialty surgery ward SS1 which has values of 0.16 and 0.08, providing some
evidence to suggest that precaution measures had a negative impact under the assumptions of the models in question (full model and no-background model).

The $\log(\beta_1 / \beta_2)$ values give very similar conclusions, and in particular posterior point values obtained by pooling estimates across all wards (using the inverse variance method as described in \cite{sutton00} (section 5.2) are all close to unity for each model. However, the posterior credible intervals for $\log(\beta_1 / \beta_2)$ indicate a high degree of posterior uncertainty for this quantity, which is most likely inherited from the high posterior variances for $\beta_1$ and $\beta_2$.

\begin{figure}[t]
\begin{center}
\includegraphics[width=1cm\textwidth,height=13cm,width=13cm,angle=0]{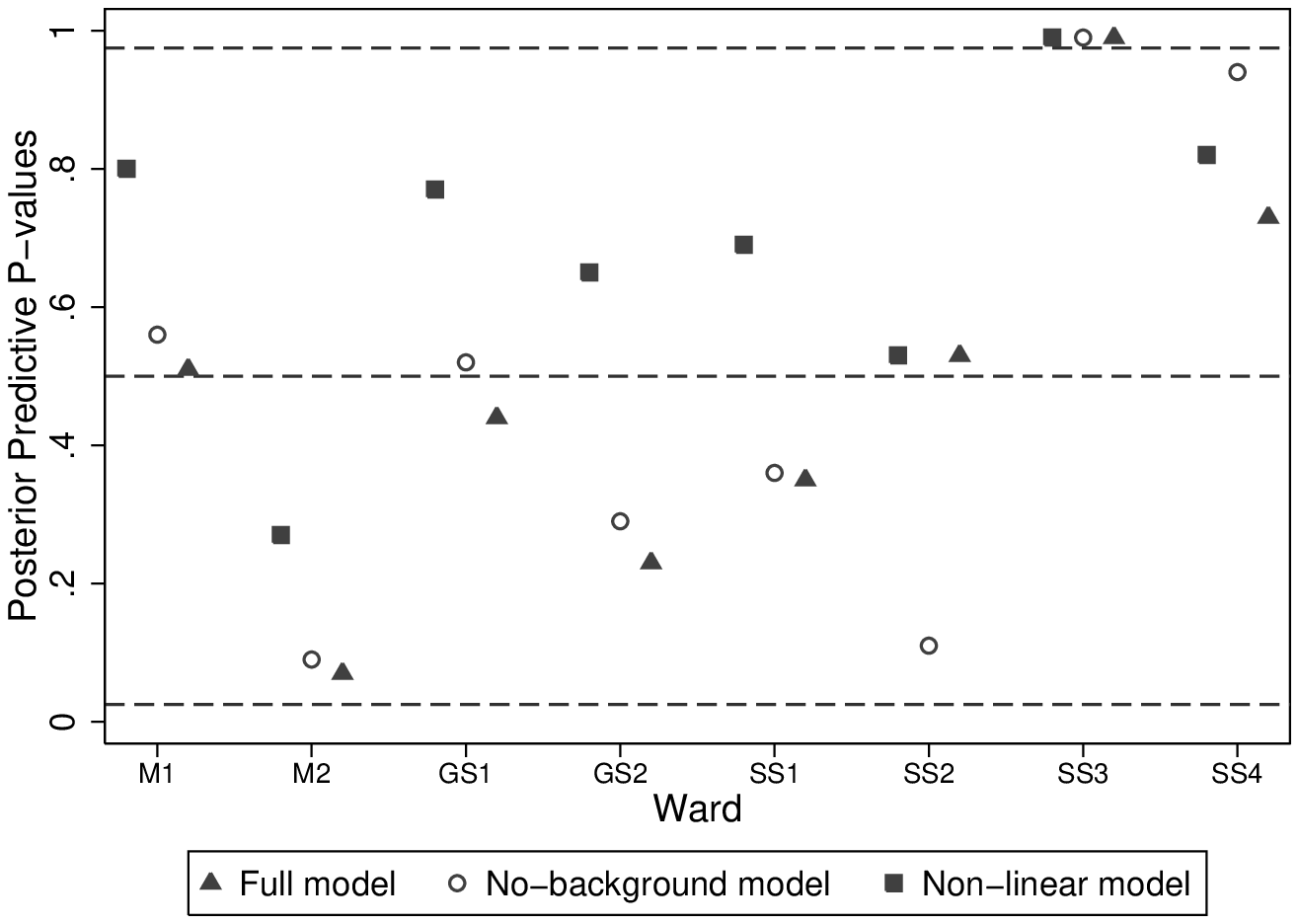}
\caption{Bayesian predictive posterior $p$-values based on number of detected colonizations. The horizontal lines represent $p = 0.025, 0.5, 0.975$ from the bottom to the top.} \label{Fig:ppp}
\end{center}
\end{figure}

{\em Model assessment.} The posterior predictive $p$-values are given in Figure \ref{Fig:ppp}. There is no strong evidence to suggest that the fitted models are inappropriate for the data, with one exception of ward SS3. Based on the $p$-values, there is no clear indication that any model is preferred over the others.

\begin{figure}[t]
\begin{center}
\includegraphics[width=1cm\textwidth,height=13cm,width=13cm, angle=-90]{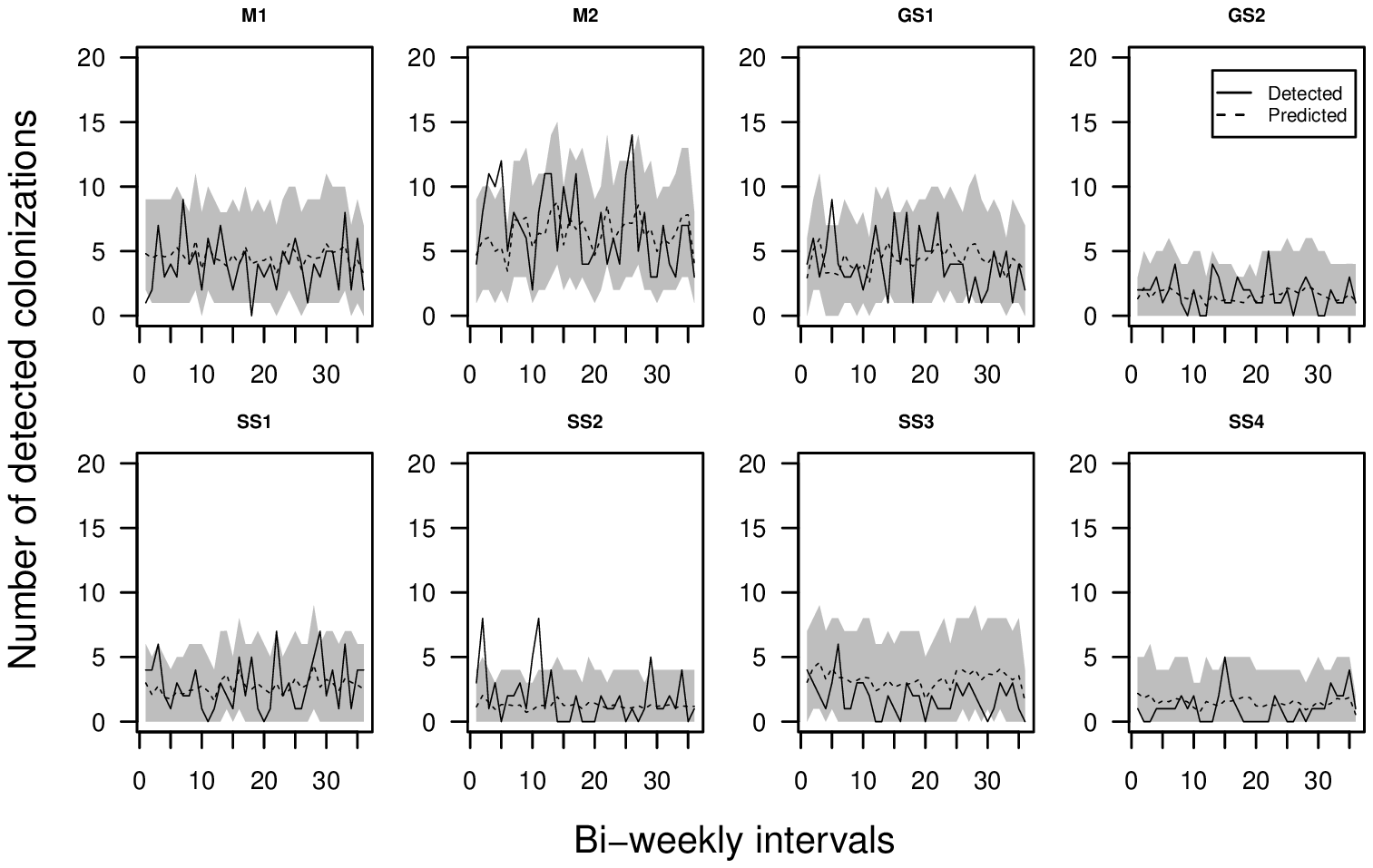}
\caption{Full model. Number of detected colonizations at 14-day intervals. Solid lines - number of detected colonizations in data; dashed lines - mean number of detected colonizations based on 2000 simulations from the predictive distribution; shaded area - 2.5$\%$ and 97.5 $\%$ percentiles from the predictive distribution.} \label{Fig:no_det_colonizations}
\end{center}
\end{figure}

\begin{figure}[t]
\begin{center}
\includegraphics[width=1cm\textwidth,height=13cm,width=13cm, angle=-90]{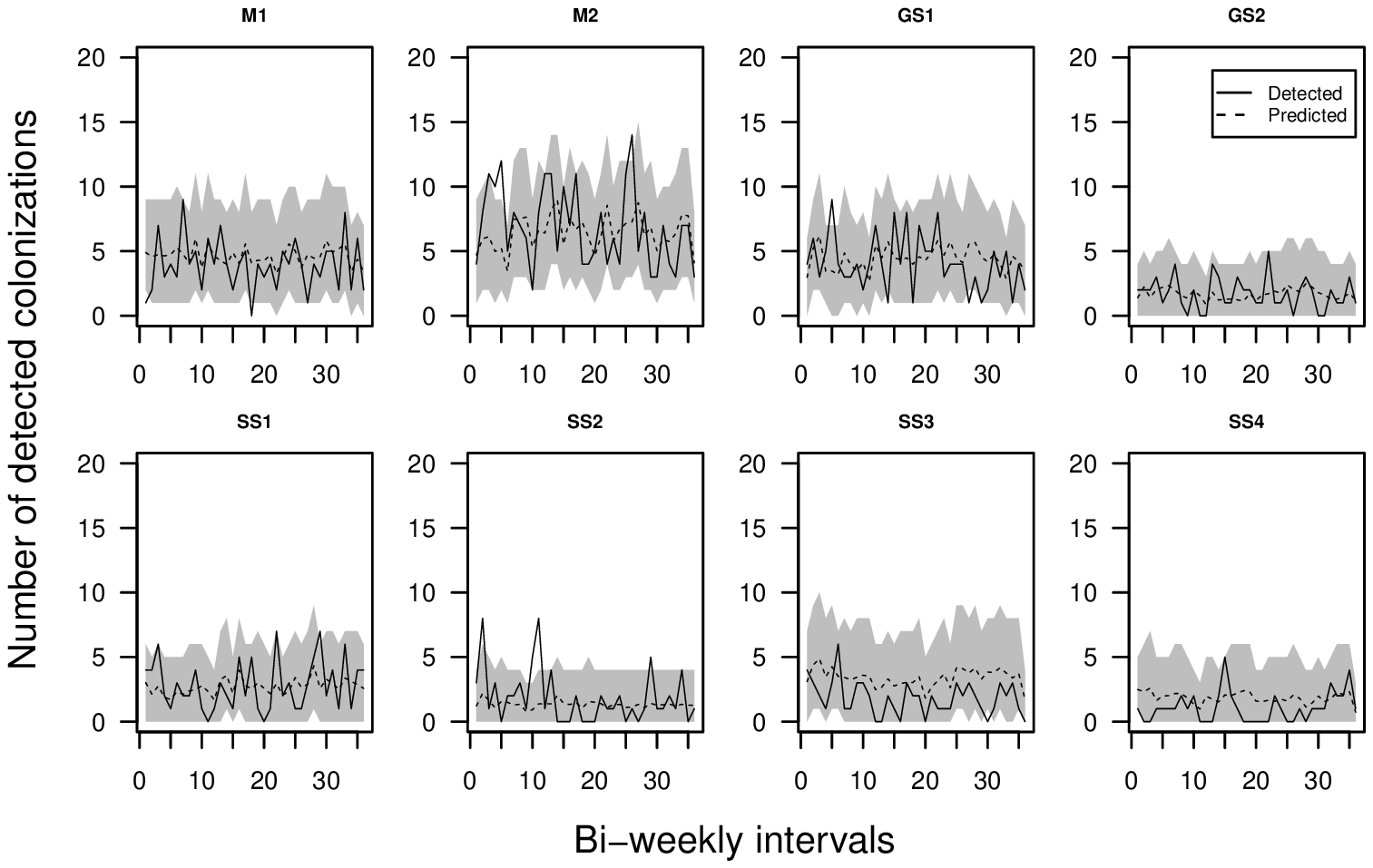}
\caption{No-background model. Number of detected colonizations at 14-day intervals. Solid lines - number of detected colonizations in data; dashed lines - mean number of detected colonizations based on 2000 simulations from the predictive distribution; shaded area - 2.5$\%$ and 97.5 $\%$ percentiles from the predictive distribution.}\label{Fig:no_det_colonizations_noback}
\end{center}
\end{figure}

\begin{figure}[t]
\begin{center}
\includegraphics[width=1cm\textwidth,height=13cm,width=13cm, angle=-90]{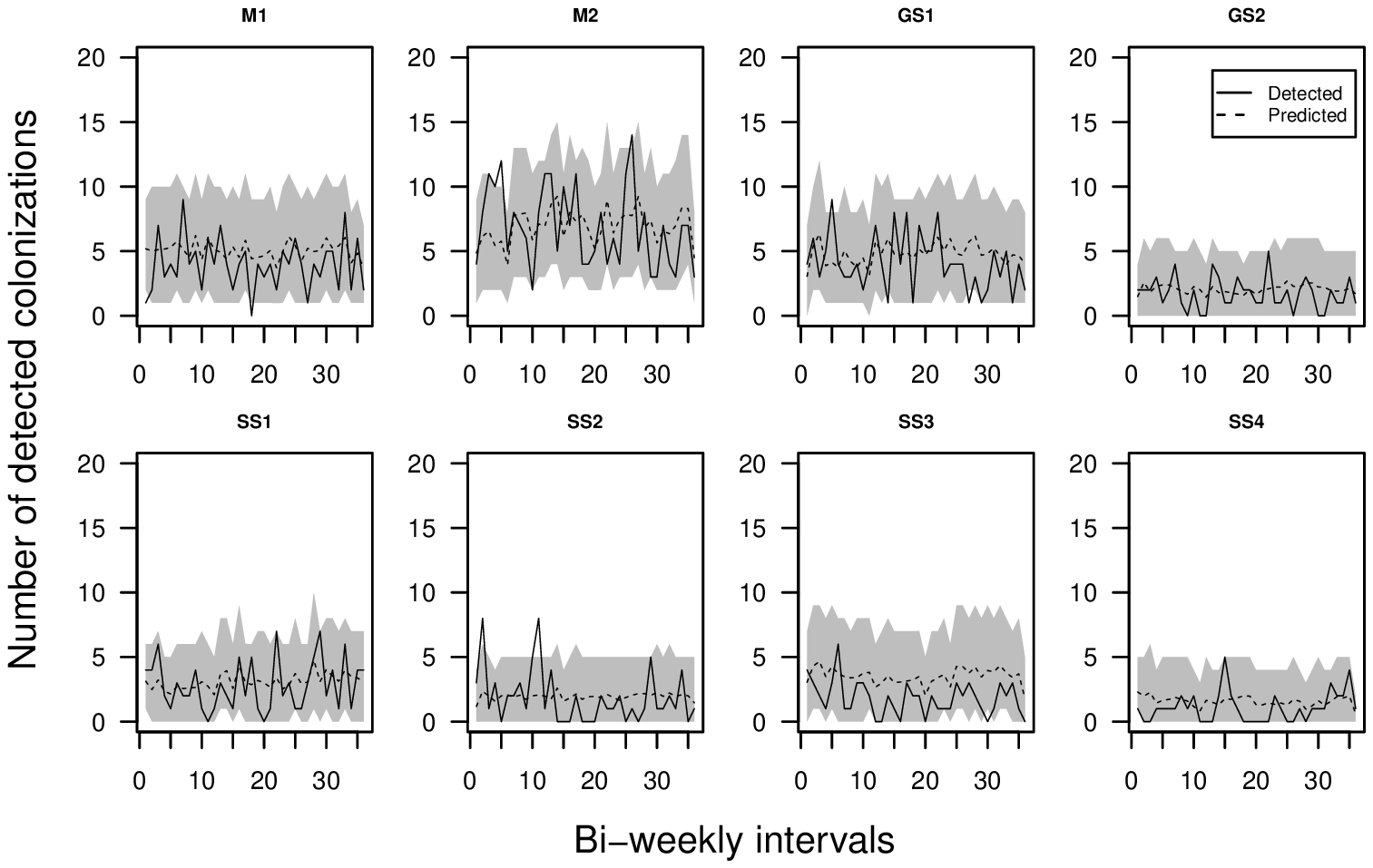}
\caption{Non-linear model. Number of detected colonizations at 14-day intervals. Solid lines - number of detected colonizations in data; dashed lines - mean number of detected colonizations based on 2000 simulations from the predictive distribution; shaded area - 2.5$\%$ and 97.5 $\%$ percentiles from the predictive distribution.}\label{Fig:no_det_colonizations_nonlinear}
\end{center}
\end{figure}

Figures \ref{Fig:no_det_colonizations}-\ref{Fig:no_det_colonizations_nonlinear} show the comparison between simulated and the observed epidemic data based on the number of detected colonizations at 14-days intervals. The figures indicate that the number of detected colonizations are reasonably well approximated by the simulated realisations, although the models can fail to capture the occasional more extreme fluctuations in the data. Overall there is no clear indication that one model is to be preferred.

{\em Model comparison.} DIC values are given in table \ref{t:DIC}.
It is notable that the no-background model is preferred in the medical and general wards, although in some cases the full model DIC values are very similar.
This suggests that the risk of new colonization events increases with the number of existing colonized patients. In contrast, for all of
the specialty surgery wards the non-linear model is favoured. This suggests that it is the presence of colonized patients, rather than
the number, that is the driving force behind new colonizations in these wards.

\begin{table}
 \vspace*{-6pt}
 \centering
 \def\~{\hphantom{0}}
  \caption{DIC values. Bold values indicate the preferred model. }
\label{t:DIC}
\begin{tabular*}{\columnwidth}{@{}l@{\extracolsep{\fill}}c@{\extracolsep{\fill}}c@{\extracolsep{\fill}}c}
\hline
        & Full model& No-background &Non-linear  \\
\hline
M1	&	1696.88	&	\textbf{1688.15}	&	1716.30 \\
M2	&	1990.56	&	\textbf{1927.34}	&	2104.64\\
GS1	&	1817.78	&	\textbf{1815.50}	&	1831.92\\
GS2	&	1260.86	&	\textbf{1196.64}	&	1343.36\\
SS1	&	948.46	&	955.94	            &	\textbf{939.00}\\
SS2	&	1186.40	&	1193.44	            &	\textbf{1171.1}\\
SS3	&	1498.58	&	1499.42	            &	\textbf{1460.34}\\
SS4	&	417.64	&	567.82	            &	\textbf{356.66}\\

\hline
\end{tabular*}\vskip18pt
\end{table}

{\em Sensitivity analysis.} In our primary analysis we have used $Exp(10^{-6})$ (mean $10^{6}$, variance $10^{12}$) as the prior distribution for each of $\beta_0, \beta_1$ and $\beta_2$. In particular, this means that the prior density for these parameters is relatively flat in the region of interest, and
so we expect the prior to have little impact on the posterior distributions. To explore this assumption, we conducted a sensitivity analysis by first using $Exp(10^{-3})$ prior distributions, and then $Exp(10^{-12})$ prior distributions. In both cases, the posterior estimates of model parameters were found to be very similar to those in the original analysis.

We have assumed that patients remained colonization for six months after they became colonized. We assessed the impact of this assumption on the posterior distribution by assuming that patients remained colonized for three months. Again, the results obtained were very similar to those in the original analysis.

\section{Discussion}
We finish by discussing the results of the analysis of the VRE data set, and then general comments on our methods.

\subsection{VRE study}

Our results show that almost half of VRE prevalence was unobserved.
This was true despite high compliance admission rectal swabs for these
ICU patients, suggesting that a single swab is insufficient for adequate
sensitivity, as has been previously suggested \cite{dagata02}.
In addition, despite the fact that weekly rectal screening for VRE was
the standard of care in these ICUs, the short average length-of-stay
meant that most patients only received one swab.
Relatively high prevalence values are not unknown; for instance \cite{bradley02} reports that about 70\% of swabs were found to be positive during a study in a UK haematology ward, while the prevalence estimated by \cite{austin99} also reaches 70\%. Furthermore, neither of these studies took into account imperfect swab test sensitivity as we have done.

For the full and non-linear models we found that the background rate of
colonization ($\beta_0$) was typically two to three times higher of that
the colonization rates attributed to colonized patients ($\beta_1, \beta_2$),
meaning that the background rate was roughly equivalent to that due to two or
three colonized patients on the ward.
Background contamination is believed to be a possible route of VRE
transmission \cite{bonten96, falk00, martinez03}
and these findings give no evidence to the contrary. However, some care is needed in interpreting $\beta_0$ within our models, since it
effectively models any colonization potential which is not associated
with colonized patients in the ward, and in particular is not restricted
to environmental contamination alone. We also found some pattern in the DIC
values in table \ref{t:DIC}, suggesting different kinds of transmission
in the specialty surgery wards compared to the other wards.

We found no evidence to suggest that the use of barrier precautions was effective
in reducing colonization. Broadly speaking, the colonization rates attributable to
patients placed under precautions were not significantly different to the
colonization rates attributable to patients who were not under precautions. One plausible
explanation is that the barrier precautions really were not discernably more effective
than standard procedures, especially given that at any one time there might only be one
or two colonized patients on the ward, making any true difference hard to detect.

\subsection{Methods}

We have described model-based methods for analysing patient-level data on nosocomial pathogens. The methods work at
an individual level by explicitly keeping track of the status of each patient and assuming a mechanism by which
patients may become colonized with the pathogen in question. One appealing aspect of this kind of modelling is that
it is often straightforward to adapt it to include additional model features, as motivated by both the available data
and the scientific questions of interest.

Our methods involved data augmentation, essentially providing estimates of quantities that are not directly observed.
Although this was primarily motivated by the fact that such missing data enable us to calculate a likelihood, the fact
that the missing data (namely, colonization times) correspond to real-life events mean that they can be exploited to
provide further information. In our application we used them to estimate quantities that describe the impact of unobserved
carriage.

We have considered three possible models for the colonization process, based on widely-used natural assumptions for how the
numbers of susceptible and colonized individuals affect the rate of new colonizations. Our purpose in doing so was to both
explore the impact of particular assumptions regarding colonization, and to see how robust our general scientific
conclusions were (e.g. regarding the efficacy of control measures) to the particular choice of model. However, it is clearly possible to
develop other models. One example is to model colonization rates in the form $\beta_i C(t)^{\delta_i}$, $i=1,2$, with
$0 \leq \delta_i  \leq 1$ parameters to estimate (see e.g. \cite{Oneill12}), so that $\delta_i=1$ gives the full model
assumption and $\delta_i=0$ gives the non-linear model assumption. However, estimating $\delta_i$ accurately would require
a substantial amount of data, particularly if there are typically only a few colonized patients on the ward at any one
time. Another option is to use something closer to non-parametric modelling, in which the colonization rate due to $k$ colonized
patients was $\beta_k$, either with or without constraints (e.g. requiring that $\beta_k$ increases with $k$).

Our approach considers each ward separately, motivated by the fact that there is no {\em a priori} reason to assume common
transmission parameters across wards. A natural way to relax this assumption would be to use a hierarchical model in which
within-ward parameters are themselves sampled from some hyper-distribution, and all the data from all wards are analysed
simultaneously, although in practice the amount of data augmentation is likely to make the MCMC algorithm infeasible.
Alternatively, if data on staff movements between wards were available, it would then be possible to model
between-ward interactions in a more explicit manner.

\section*{Acknowledgements}

This research was funded by a Welcome Trust grant (number 076850, PI: P.D. O'Neill). Data collection was funded by the CDC Prevention Epicenters Program and the National Institutes of Health (K23AI64161, S.S. Huang).
BSC was supported by The Medical Research Council and Department for International Development
(grant number MR/K006924/1) and works within the Wellcome Trust Major Overseas Programme in SE Asia
(grant number 106698/Z/14/Z).

We thank the reviewers for their helpful suggestions.

\end{document}